# Tutorial: Analog Matrix Computing (AMC) with Crosspoint Resistive Memory Arrays

Zhong Sun, *Member, IEEE*, Daniele Ielmini, *Fellow, IEEE*

*Abstract*—Matrix computation is ubiquitous in modern scientific and engineering fields. Due to the high computational complexity in conventional digital computers, matrix computation represents a heavy workload in many data-intensive applications, *e.g.*, machine learning, scientific computing, and wireless communications. For fast, efficient matrix computations, analog computing with resistive memory arrays has been proven to be a promising solution. In this Tutorial, we present analog matrix computing (AMC) circuits based on crosspoint resistive memory arrays. AMC circuits are able to carry out basic matrix computations, including matrix multiplication, matrix inversion, pseudoinverse and eigenvector computation, all with one single operation. We describe the main design principles of the AMC circuits, such as local/global or negative/positive feedback configurations, with/without external inputs. Mapping strategies for matrices containing negative values will be presented. The underlying requirements for circuit stability will be described via the transfer function analysis, which also defines time complexity of the circuits towards steady-state results. Lastly, typical applications, challenges, and future trends of AMC circuits will be discussed.

## I. INTRODUCTION

IN the era of big data, the demand of low-latency, high-efficiency information processing challenges conventional digital computers, whose performance, however, has been limited by the slowdown of Moore's law and the intrinsic communication bottleneck in the von Neumann architecture [1], [2]. Many of the applications consist of intensive matrix computations, *e.g.*, scientific computing, webpage searching, and product recommendation in the cloud, and pattern recognition, MIMO communication at the edge. Matrix computing and especially matrix equation solving, however, is usually complicated, causing severe issues in terms of latency and energy dissipation.

In recent years, alternative computing concepts have been intensively investigated to accelerate matrix computations. Among them, analog computing with resistive memory is a highly promising paradigm. After a long period of abandonment, analog computing has recently gained renewed attention, thanks to its fast response speed and the embedded massive parallelism. In contrast to the early analog computers which generally dealt with differential equations, the new analog computing concept focuses on matrix problems, aiming at the acceleration of both cloud and edge applications [3]. Schemes based on both conventional CMOS and emerging device technology have been developed for analog matrix computing (AMC). In the latter case, resistive random-access memory (RRAM) has been considered as an excellent candidate. Compared to the fully-CMOS approach [4]-[6], AMC circuits based on crosspoint RRAM arrays [7]-[9] are much more area-efficient, thanks to the compact structure and the analog conductance capability of RRAM devices [10]. Additionally, RRAM-based AMC implies inherently the in-memory computing concept, thus eliminating the infamous von Neumann bottleneck in conventional computers.

In this Tutorial, we introduce RRAM-based AMC circuits for several basic matrix computations, namely matrix multiplication, inversion, pseudoinverse and eigenvector. The circuits were designed according to the mathematical expressions of matrix problems to realize analog solutions in real time. The various circuit topologies are presented for each computing function. For matrix computations with negative entries, possible implementation schemes are presented. The fundamental stability requirements of the circuits are summarized to support unambiguously their feasibility for practical applications and the corresponding time complexities. The typical applications and main challenges of AMC circuits are discussed. Finally, the Tutorial is concluded with a summary of the main features of the AMC circuits.

## II. RRAM-BASED ANALOG MATRIX COMPUTING CIRCUITS

RRAM is a two-terminal nonvolatile memory device whose resistance (or conductance) can be changed by applying external voltage pulses. According to the electrical stimuli, ionic migration results in different structural configurations, hence different conductance states [10]. Beyond the conventional binary memory applications, RRAM can show multi-level states for analog value storage. Analog RRAM has been shown to be capable of 64 conductance states

This work was supported in part by the National Key Research and Development Program of China under Grant 2020YFB2206001, in part by the National Natural Science Foundation of China (NSFC) under Grant 62004002, Grant 92064004 and Grant 61927901, and in part by the 111 Project under Grant B18001. *(Corresponding authors: Zhong Sun; Daniele Ielmini.)*

Zhong Sun is with the Institute for Artificial Intelligence and the School of Integrated Circuits, Peking University, Beijing 100871, China (e-mail: zhong.sun@pku.edu.cn).
Daniele Ielmini is with the Dipartimento di Elettronica, Informazione e Bioingegneria, Politecnico di Milano, 20133 Milan, Italy (e-mail: daniele.ielmini@polimi.it)

(equivalently 6 bits) or more [11]-[13]. Also, dedicated program/verify techniques may be used to achieve an arbitrary conductance state with a pre-defined error window [14]. RRAM devices are usually organized as crosspoint arrays, thus achieving the highest storage density in the plane. In addition, the crosspoint array forms a physical matrix where the RRAM conductance values represent the individual matrix entries. Based on the crosspoint architecture, multiple matrix operations can be realized by a proper configuration of the AMC circuits.

### A. Matrix Multiplication Circuit

The most straightforward application of crosspoint RRAM array is for the acceleration of matrix-vector multiplication (MVM). As shown in Fig. 1(a), matrix $A$ is mapped as conductance values of RRAM devices in the crosspoint array with reference to a unit conductance value $G_0$. Voltage of a vector $x$ are applied to the columns of the array, while the current is collected from each row. Usually, a transimpedance amplifier (TIA) is used to convert the row current to an output voltage [7]. Due to the virtual ground realized for input terminals of TIAs, the circuit output forms the MVM vector $y$, namely:

$$y = Ax, \qquad (1)$$

where $A$ has an arbitrary size $n \times m$, $x$ size is $m \times 1$, and $y$ size is $n \times 1$. As there is a sign inversion in the conversion operation due to the negative voltage gain of the TIA, the input vector $x$ is inverted during mapping as shown in the figure. In the hardware implementation, a RRAM cell may consist of a passive 1R device or a one-transistor/one-resistor (1T1R) structure [14]-[16]. In the latter case, an additional set of wordlines is needed to properly bias the transistor gates during the MVM operation. A sensing amplifier (SA) may also be used for current readout. SAs are widely used in memory architectures and applies conveniently to the more mature binary RRAM technology [17].

Since the device conductance can only be positive, a matrix containing negative entries has to be split as $A = A_+ - A_-$,

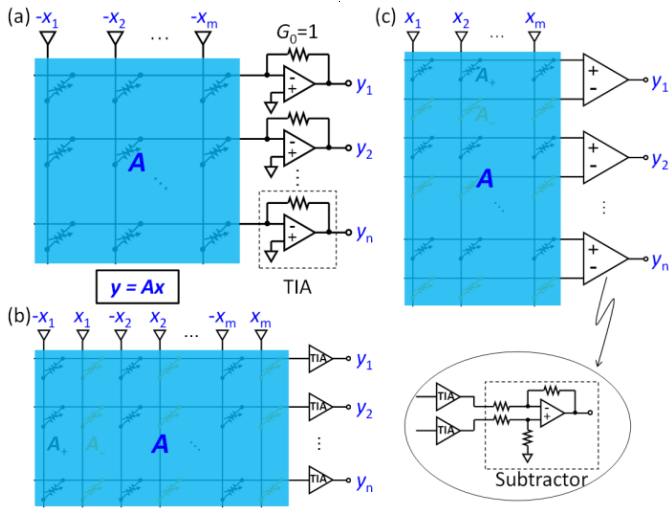

Fig. 1. AMC circuits for MVM computation. (a) AMC circuit for MVM of a positive matrix $A$. (b) Column-wise splitting AMC circuit for MVM with negative entries. (c) Row-wise splitting AMC circuit, where the inset shows a possible implementation of readout component.

where $A_+$ and $A_-$ both contain only positive entries. Two types of splitting methods may be used, namely column-wise and row-wise splitting, as shown in Fig. 1(b) and 1(c), respectively. In the former case, each column of matrix $A$ is represented by the conductance difference of two columns of crosspoint RRAM devices that map $A_+$ and $A_-$, which in turn are applied respectively with voltages that represent an element of vector $-x$ and $x$. TIAs are used for current readout as in the positive matrix case. In row-wise splitting, each row of $A$ is mapped by two rows of RRAM devices, which connect to two terminals of one readout component. The input vector $x$ is directly applied as voltage bias of the array columns. The inset of Fig. 1(c) shows a possible implementation of the readout circuit, which uses three operational amplifiers (OAs) [15]. Optimized methods have been developed to simplify this component, using 2 or less OAs for each [18], [19].

Once all the input voltages are applied to the AMC circuit of Fig. 1, the product results are obtained simultaneously in parallel, which is fundamentally different from the sequential data processing in digital computers. Circuit analysis reveals that the computing time to reach the steady-state output is linearly dependent on the maximal row sum of the mapped conductance matrix [20]. As a result, depending on the matrix structure, the time complexity of MVM with AMC may be different, e.g., $O(n^{1/2})$ or $O(\log n)$. However, $O(1)$ complexity can be easily achieved by adjusting the feedback conductance of TIAs according to the matrix size, which requires a pre-processing step with small overhead.

### B. Matrix Inversion Circuit

The inverse problem of MVM in Eq. (1) consists of solving the following matrix equation:

$$Ax = y, \qquad (2)$$

where $y$ is known while $x$ is unknown. Eq. (2) represents a system of linear equations, which has a higher complexity compared to MVM in digital computers. Fig. 2 shows an AMC circuit, where feedback loops are configured between rows and columns of a crosspoint RRAM array via OAs to carry out matrix inversion [8]. Each row in the crosspoint array is connected to the inverting input terminal of an OA, while columns are connected to OA outputs. During the circuit operation, the row potential is forced to virtual ground by the negative feedback. The input vector $y$ is represented by external voltages applied to load resistors. The output voltage represents vector $x$, which generates MVM currents at the row lines of virtual ground. According to the Kirchhoff's current law, the currents at the array rows should be zero, thus $Ax - y = 0$. Since $A$ and $y$ are given, $x$ provides the solution to Eq. (2), namely:

$$x = A^{-1}y, \qquad (3)$$

where $A^{-1}$ is the inverse matrix of $A$. For matrix inversion, $A$ must be square ($n \times n$) and non-singular. It is interesting to note that the MVM circuit in Fig. 1(a) and the matrix inversion circuit in Fig. 2 have the same components, however, the different connection topologies result in the exactly inverse functions. Note also that the OAs in Fig. 2 operate with closed loops where the crosspoint RRAM devices play the role of feedback network, thus enabling the negative feedback mechanism.

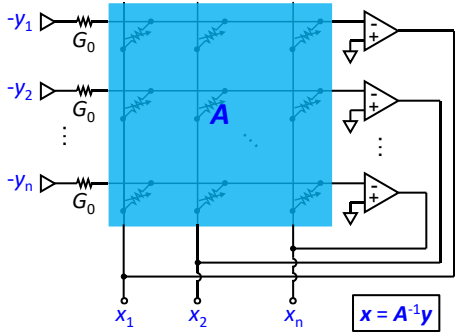

Fig. 2. AMC circuit for matrix inversion, namely solving a linear system $Ax = y$. Basically, this circuit is a matrix inversion operator $A^{-1}$. If matrix is positive definite, the circuit is stable and gives correct solutions.

The stability of the circuit can be understood by studying the transfer function [21]. The transfer function in Fig. 2 can be described by the matrix equation $-U^{-1}[Ax(s) - y(s)]L(s) = x(s)$, where $s$ is the complex frequency, $L(s)$ is the open-loop gain of the OA and $U$ is a diagonal matrix defined as $U = diag(1 + \sum_j A_{1j}, 1 + \sum_j A_{2j}, \cdots, 1 + \sum_j A_{nj})$. Consider the single-pole transfer function for the OAs [22], namely $L(s) = \frac{L_0}{1+s/\omega_0}$, where $L_0$ is the DC open-loop gain, $\omega_0$ is the 3-dB bandwidth, and their product defines the gain bandwidth product (GBWP) of the OA, i.e., $f_{GBWP} = L_0 \omega_0$. As a result, the transfer function of the circuit is obtained, namely,

$$x(s) = \left(A + \frac{s}{f_{GBWP}} U\right)^{-1} y(s). \quad (4)$$

According to Eq. (4), the poles of the circuit are given by $s = -f_{GBWP} \cdot \lambda(U^{-1}A)$, where $\lambda(U^{-1}A)$ are the eigenvalues of matrix $U^{-1}A$. To guarantee the circuit stability, all poles should be in the left half plane, thus all eigenvalues of matrix $U^{-1}A$ should have positive real parts. Such a condition can be perfectly satisfied by assuming that $A$ is positive definite (PD). In this case, the matrix $U^{-1}A$ has the same eigenvalues as $A^{\frac{1}{2}}U^{-1}A^{\frac{1}{2}}$, while the latter is positive definite. Therefore, eigenvalues of $U^{-1}A$ are all real and strictly positive.

Since circuit response is ultimately determined by the dominant pole, computing time of the circuit is controlled by the minimal eigenvalue ($\lambda_{min}$) of matrix $U^{-1}A$. Theoretically, the upper bound of computing time is proportional to $\frac{1}{\lambda_{min}}$, which suggests that a larger $\lambda_{min}$ of $U^{-1}A$ results in a faster computing speed. To illustrate the effect of $\lambda_{min}$, Fig. 3 compares the transient responses for solving two different linear systems of PD matrices $A_1$ and $A_2$. Fig. 3(a) shows the poles of the corresponding two circuits obtained from $\lambda(U^{-1}A)$ and normalized by $f_{GBWP}$, all on the real axis and in the left half plane for both cases. Specifically, $\lambda_{min}$ is 0.03 and 0.007 for $A_1$ and $A_2$, respectively. Fig. 3(b) shows the transient response of the two circuits, indicating that the response time in the former case is 1.6 µs, while the latter is 4.2 µs, demonstrating the key role of $\lambda_{min}$ for determining circuit speed. Depending on the specific matrix structure, various time complexities dictated by $\lambda_{min}$ may be presented, e.g., $O(\log n)$ or $O(1)$ [21]. Also, the computing time is fundamentally limited by $f_{GBWP}$, the optimization of which would speed up the solution.

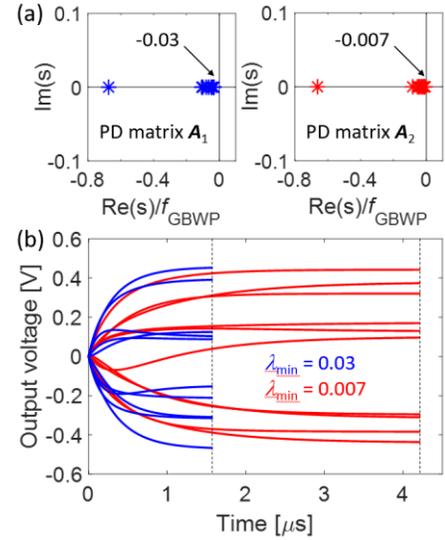

Fig. 3. Computing time of the matrix inversion circuit. Two $10 \times 10$ PD matrices $A_1$ and $A_2$ are considered. (a) Poles of the circuit based on eigenvalues of matrix $U^{-1}A$. The dominant poles are labeled. (b) Transient curves of output voltages of the circuit for solving the two matrix inversion problems. Response time of the circuit is defined as the moment $t$ when $\|x(t) - x^*\|_2$ is less than 1%, where $x^*$ is the steady-state result.

For inversion of a matrix that contains negative entries, the original matrix $A$ has to be split as in the MVM case [23]. Currently, only the column-wise splitting scheme using analog inverters has been proposed, as shown in Fig. 4(a). The circuit is stable if $A$ is PD and it is split as $A_+ = \frac{|A|+A}{2}$ and $A_- = \frac{|A|-A}{2}$. Due to the two sets of OAs, this circuit is described as a quadratic eigenvalue problem (QEP) [24], which yields $2n$ poles. In this case, all poles of the circuit are in the left half plane, however, they are not all real, as shown by the example in Fig. 4(b). Fig. 4(c) shows the corresponding transient response of the circuit, indicating a computing time of 12.08 µs. The dominant pole can be calculated through the minimal (real part of) eigenvalue of a $2n \times 2n$ matrix based on matrices $A_+$ and $A_-$.

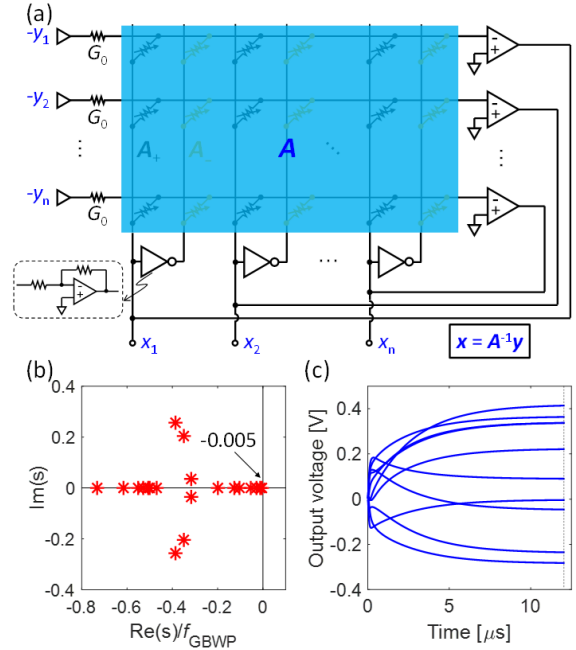

Fig. 4. (a) Matrix inversion circuit for matrices containing negative entries. Inset shows the structure of an analog inverter, where the OA is same as the OAs connected to rows. (b) Poles of the circuit for a $10 \times 10$ PD matrix, where some poles show imaginary parts. (c) Transient output voltages of solving a $10 \times 10$ matrix inversion problem.

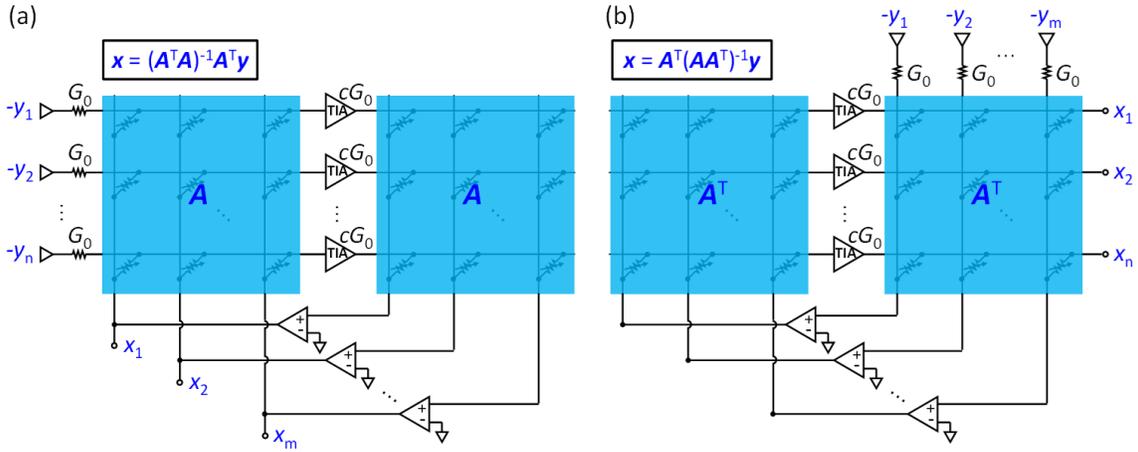

Fig. 5. AMC circuits for pseudoinverse computation. (a) Left inverse, and (b) Right inverse circuits for solving overdetermined or underdetermined linear systems. In both circuits, each crosspoint RRAM array has $n$ rows and $m$ columns ($n > m$), representing $A$ and $A^T$ in (a) and (b), respectively.

### C. Matrix Pseudoinverse Circuit

While matrix inversion deals with only square matrices, non-square matrices are frequently encountered in many important algorithms, such as linear/logistic regression, compressive sensing and so on. We have extended the AMC inversion circuit to non-square matrices to compute the pseudoinverse of matrix [9]. For an overdetermined linear system with more equations than unknowns, i.e., $A$ in Eq. (2) is a tall matrix ($n \times m$ with $n > m$), the problem is unsolvable. It can only be solved in the sense of minimizing the squared error by the concept of pseudoinverse (or Moore-Penrose inverse [25]), that is

$$x = A^+ y = (A^T A)^{-1} A^T y, \quad (5)$$

where $A^+$ and $A^T$ denote pseudoinverse and transpose of matrix $A$, respectively.

The AMC circuit for obtaining the solution of Eq. (5) is shown in Fig. 5(a). In the circuit, two crosspoint RRAM arrays mapping the same matrix $A$ are connected through a set of TIAs and another set of OAs. The TIA feedback conductance $c$ does not affect the steady-state output, rather it controls the circuit dynamics. OAs work in the positive gain mode, thus facilitating an overall negative feedback mechanism of the circuit. Input currents representing vector $y$ are supplied to the input terminals of the TIAs, while the currents collected at columns of the right array must all be zero, thus resulting in the matrix equation $A^T(y - Ax)/c = 0$. After a minor rearrangement, this equation leads to Eq. (5), which is thus automatically solved by the AMC circuit. The circuit applies to linear regression, which is a typical problem of solving overdetermined linear systems, and can be extended to logistic regression, which is particularly useful for the training of the classification layer in deep neural networks.

The pseudoinverse in Eq. (5) is known as left inverse [26]. Other matrix problems such as compressive sensing and linear programming are solved by underdetermined linear systems, where the number of equations is smaller than the number of unknowns. In such cases, $A$ in Eq. (2) is a broad matrix ($m \times n$ with $n > m$). Consequently, the right inverse has to be used to obtain the solution as follows:

$$x = A^+ y = A^T (A A^T)^{-1} y. \quad (6)$$

The AMC circuit for obtaining the right inverse solution is shown in Fig. 5(b). In contrast to the left inverse circuit, input currents are provided to the OAs in the right inverse circuit, and the transpose of $A$ is mapped in the two crosspoint arrays. The feedback loop of two circuits are exactly the same, thus featuring the same underlying stability mechanism. The rigorous stability study results in a QEP for describing the poles of the circuit [27]. Specifically, the circuit has $n + m$ poles in the left half plane and $n - m$ poles at the origin, thus is always stable in the sense of Lyapunov [28]. The response time of the circuit is determined by the dominant pole, resulting in an unconventional time complexity where the computing time may even decrease as the problem size increases. Parameters including feedback conductance $c$ of TIAs and GBWPs of the two sets of OAs in the circuit can be optimized synergistically to improve the performance in terms of speed, signal-to-noise ratio, and power consumption. Pseudoinverse problems for matrices containing negative values may be split as Fig. 4(a), which inevitably makes the circuit more complicated. Fortunately, for linear regression problems, where the matrix $A$ has one column of ones, all data can be shifted to be positive and solved by the circuit in Fig. 5(a).

Both left and right inverse circuits can be also used to solve square matrix inversion problems expressed by Eq. (2). Thanks to their intrinsic stability, the circuits can solve any linear system irrespective of the matrix $A$, at the expense of one more memory array than the single-array circuit of Fig. 2. For this reason, the circuits in Fig. 5 represent a universal solution to any linear systems regardless of the matrix dimensions.

### D. Matrix Eigenvector Circuit

The global feedback concept allows the computation of

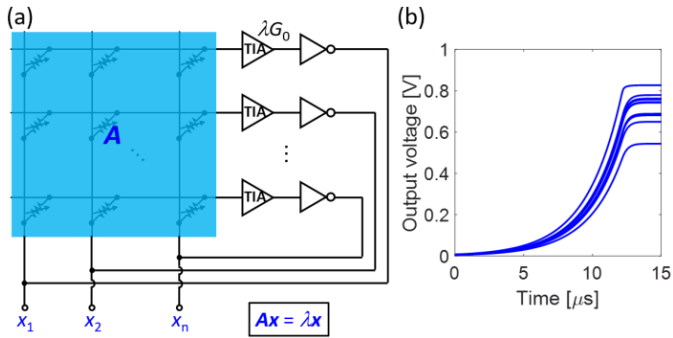

Fig. 6. (a) AMC circuit for computing eigenvector of positive eigenvalue $\lambda$. (b) Transient output voltages of solving a $10 \times 10$ matrix eigenvector problem.

matrix eigenvectors [8], namely the vector $x$ in the equation
$$Ax = \lambda x, \qquad (7)$$
where $\lambda$ is an eigenvalue of matrix $A$. The circuit for a positive $\lambda$ is shown in Fig. 6(a), where $\lambda$ is mapped as feedback conductance of the TIAs and analog inverters are connected in series to the TIAs to invert the sign of the output voltage. In the circuit, the output voltage at the analog inverters is subject to the equation $-(-Ax/\lambda) = x$, which suggests the solution to the eigenvector equation of Eq. (7). For a negative eigenvalue $\lambda$, its absolute value should be mapped in the TIAs, while the inverters are removed.

A summary of design principles of the AMC circuits is presented in Table I. While the matrix multiplication, inversion, pseudoinverse circuits work with negative feedback mechanism (local or global), the eigenvector circuit has a positive feedback loop. Also, the former three circuits have external input sources, while the eigenvector circuit works in a self-sustained mode. Voltage signals in the circuit are amplified by TIAs and inverters in continuous time, resulting in steady-state outputs proportional to eigenvector elements, where the largest voltage (in magnitude) reaches the saturation level of its OA [29]. The physical amplification process is equivalent to the power iteration algorithm in discrete time (Fig. 6(b)), showing a similar feature that only the dominant eigenvector (corresponding to the positive or negative largest eigenvalue) can be solved by the circuit. Such a self-sustained mechanism requires that the loop gain of the circuit is higher than 1. To satisfy this condition, the mapped eigenvalue has to be slightly less than the nominal one, which in turn affects the computing accuracy [29]. Theoretical analysis shows that computing time of the eigenvector circuit is solely determined by the deviation introduced during eigenvalue mapping, suggesting a tradeoff between accuracy and speed. These results demonstrate the $O(1)$ time complexity of the AMC eigenvector circuit.

### III. Discussion

As an emerging computing paradigm, AMC offers extremely fast solutions to basic matrix computations, which is desired in many important scenarios requiring low latency and high energy efficiency. For instance, solving linear systems is the core of modern scientific computing, the matrix inversion circuit has been considered for solving large-scale problems, combined with other linear algebra techniques such as preconditioning and domain decomposition [30]-[32]. Table II shows the performance metrics of AMC circuits for typical applications. Note that, although the performance evaluation relies heavily on applications and datasets, the AMC circuits generally show orders of magnitude of higher energy efficiency.

TABLE I. Design principles of AMC circuits.

| Design principles | AMC | Matrix multiplication | Matrix inversion | Pseudo-inverse | Eigenvector |
|---|---|---|---|---|---|
| Feedback | | Local | Global | Global | Global |
| | | Negative | Negative | Negative | Positive |
| External input | | w/ | w/ | w/ | w/o |

To achieve reliable AMC in practical applications, several challenges across the device, circuit, architecture levels have to be addressed. Regarding RRAM devices, a major obstacle for analog computing is the statistical conductance variation. Iterative verify algorithm may be adopted, at the expense of long programming time [34]. Device retention is another aspect, as the conductance fluctuation may result in accuracy degradation, which is particularly undesired in precision-sensitive tasks. As a compromise for device reliability and operational ease, RRAM with lower bit precision might be adopted, combined with high-level architecture/algorithm innovations. For 1R structure, sneak current path is a severe problem if only a fraction of the rows/columns in the array are involved for computing. It can be overcome through grounding the unselected crosspoint rows/columns, thanks to the fact that the selected crosspoint rows are all virtual ground in the AMC circuits. By contrast, it is more convenient with 1T1R structure to select sub-arrays that are free of sneak current disturbance.

Parasitic resistance/capacitance in the crosspoint RRAM arrays is a main issue in AMC circuits, which causes accuracy degradation and response delay. Input and output wire resistances have a similar impact, especially for the global feedback circuits, thus limiting the feasible size of memory array (or matrix). Also, analog noise, area occupation, power consumption and bandwidth of OAs may significantly limit accuracy, latency and efficiency of the circuits. Therefore, efforts to design optimized OAs would be valuable. Due to these non-ideal properties of devices and circuits, the scalability of the AMC architecture is an important challenge. As a result, strategies for developing scalable AMC is of fundamental importance. Typical solutions may include bit slicing [35] and submatrix division. On the other hand, as all AMC circuits work with the same core components, *i.e.*, crosspoint RRAM arrays and OAs, efforts to develop reconfigurable architectures that are capable of different matrix operations would increase the AMC universality and hardware efficiency.

TABLE II. Performance metrics of AMC circuits.

| Perf. AMC | Circuit area | Application (Dataset) | Accuracy | Throughput (equivalent) | Energy efficiency | Ref. |
|---|---|---|---|---|---|---|
| Matrix multiplication | 0.064 mm² | Neural network inference (MNIST) | 96.2% | 0.082 TOPs | 11 TOPs/W (>100x GPU) | [14] |
| Matrix inversion | 0.543 mm² | Linear solver preconditioner (SuiteSparse matrix collection) | 4-bit Approx. | - | 1025x CPU | [31] |
| Pseudoinverse | - | Neural network learning (MNIST) | 94.2% | 16.1 TOPs | 45.3 TOPs/W (19.7x TPU) | [9] |
| Eigenvector | - | PageRank (Harvard500) | 95% | 0.183 TOPs | 362 TOPs/W (157x TPU) | [33] |

## IV. Conclusion and Perspectives

In this Tutorial, we introduce AMC circuits based on crosspoint resistive memory arrays and feedback connection with OAs. Four AMC operator circuits are described, namely (i) matrix multiplication, (ii) matrix inversion, (iii) matrix pseudoinverse and (iv) matrix eigenvector. All circuits except for the matrix inversion circuit are stable for any given matrix, while the latter fits perfectly to PD matrix. The computing time of all circuits except for the eigenvector circuit is controlled by the minimal eigenvalue (or real part of eigenvalue) of an associated matrix, showing unconventional time complexities (depending on matrix structure), while the latter is validly $O(1)$ regardless of the specific matrix structure. While these AMC circuits can address solving linear problems, it would be desirable to design similar circuits for solving nonlinear matrix problems in the same framework, such as compressive sensing, optimization or programming. The combination of both linear and nonlinear matrix problem solving will extensively enrich the range of AMC applications.